\documentclass[prb,preprint,showpacs,superscriptaddress]{revtex4}

\usepackage{amsmath,color}
\usepackage{epsfig,graphicx}
\usepackage{rotating}

\usepackage{eso-pic}
\usepackage{type1cm}
\begin{document}
\title{Rotational dynamics and polymerization of C$_{60}$ in C$_{60}$-cubane crystals: A molecular dynamics study}

\author{V. R. Coluci}\email[\footnotesize{Author to whom correspondence should be addressed. FAX:+55-19-21133364. Electronic address: }]{vitor@ceset.unicamp.br}

\affiliation{Center for High Education on Technology, 
             University of Campinas - UNICAMP 13484-332, Limeira,
             SP, Brazil}

\affiliation{Applied Physics Department, Institute of Physics P.O.Box 6165,
             University of Campinas - UNICAMP 13083-970, Campinas,
             SP, Brazil}

\author{F. Sato}
\affiliation{Applied Physics Department, Institute of Physics P.O.Box 6165,
             University of Campinas - UNICAMP 13083-970, Campinas,
             SP, Brazil}

\author{S. F. Braga}
\affiliation{Applied Physics Department, Institute of Physics P.O.Box 6165,
             University of Campinas - UNICAMP 13083-970, Campinas,
             SP, Brazil}

\author{M. S. Skaf}
\affiliation{Institute of Chemistry P.O.Box 6154,
             University of Campinas - UNICAMP 13084-862, Campinas,
             SP, Brazil}

\author{D. S. Galv\~ao}
\affiliation{Applied Physics Department, Institute of Physics P.O.Box 6165,
             University of Campinas - UNICAMP 13083-970, Campinas,
             SP, Brazil}

\date{\today}

\begin{abstract}
We report classical and tight-binding molecular dynamics simulations of the C$_{60}$ fullerene and cubane molecular crystal in order to investigate intermolecular dynamics and polymerization processes. Our results show that, for 200 K and 400 K, cubane molecules remain basically fixed, presenting only thermal vibrations, while C$_{60}$ fullerenes show  rotational motions. Fullerenes perform  ``free'' rotational motions at short times ($\lesssim$ 1 ps), small amplitude hindered rotational motions (librations) at intermediate times, and rotational diffusive dynamics at long times ($\gtrsim$ 10 ps). The mechanisms underlying these dynamics are presented. Random copolymerization among cubanes and fullerenes were observed when temperature is increased, leading to the formation of a disordered structure. Changes in the radial distribution function and electronic density of states indicate the coexistence of amorphous and crystalline phases. The different conformational phases that cubanes and fullerenes undergo during the copolymerization process are discussed.
\end{abstract}



\maketitle

\section{Introduction}
The discovery of C$_{60}$ fullerenes~\cite{1} opened a new field in theoretical and experimental research of carbon based materials \cite{livro}. At room temperature these molecules crystallize in a face-centered-cubic (fcc) solid structure \cite{cox} exhibiting semiconducting behavior (bandgap of about 1.5 eV \cite{saito}). When the crystals are either hydrostatically compressed \cite{sun,iwasa,regueiro,Goze,Jansson} or exposed to visible or ultra-violet light sources \cite{Rao}, a polymerization process of the C$_{60}$ molecules is observed. Due to the large size of the C$_{60}$ molecules, the interstitial cavities in the crystalline C$_{60}$ can accommodate various guest species. Attempts were made to intercalate C$_{60}$ crystals with electron donors (and other atoms) in order to vary their electronic properties. Results of doping with alkaline metals lead to the appearance of exotic superconductivity with critical temperatures around 30 K or larger \cite{Ranjan,Stenger,JPCS}. Other atoms and molecules, such as rare gases \cite{gas1,gas2} and molecular oxygen \cite{Russian} were also used as dopants. 

Recently, heteromolecular crystals of C$_{60}$ fullerene and cubane (C$_8$H$_8$) \cite{cubano} have been prepared from aromatic solutions by evaporating solvent and adding isopropyl alcohol as precipitant \cite{pekker1}. These crystals (Fig.~1) present an interesting phase diagram showing orientational-ordering phase transitions. At atmospheric pressures and temperatures below 140 K, the crystal exhibits an orthorhombic symmetry. From 140 K up to 470 K, the C$_{60}$ molecules are located on the lattice sites of an fcc crystal with the cubane molecules on the octahedral voids. In this temperature range, the C$_{60}$ molecules are free to rotate whereas cubanes behave like static bearings in a so-called rotor-stator phase \cite{pekker1}. Above 470 K, solid-state reactions take place among cubanes and fullerenes leading to different crystalline and, eventually, amorphous phases due to copolymerization \cite{pekker2a}. Similar phase diagrams are also observed for other fullerenes, such as C$_{70}$ and C$_{84}$ \cite{pekker2}. 

Despite experimental studies about the polymerization process that takes place during high temperature treatments \cite{pekker2,pekker2a} and under pressure \cite{pekker3}, providing information of the possible conformations of the resulting polymer, the molecular aspects involved in these processes are still largely unknown. For instance, experimental results suggest that the resulting compound after thermal treatment (470 K) is a copolymer of C$_{60}$ with a decomposition product of cubane (dihydropentalene) \cite{pekker2}. However, other possibilities for the copolymer can not be ruled out even when other decomposition products of cubane are considered. In this case, similar copolymers could also be formed leading to different properties. While detailed information on such processes is not easily accessible from experiments, atomistic simulations can be a very important tool to provide additional insights and molecular level understanding of the experimental data. In order to investigate the dynamical processes associated with C$_{60}$-cubane crystals we present here atomistic simulations of the C$_{60}$-cubane heteromolecular crystal. 

Firstly, we have analyzed the C$_{60}$-cubane crystal in its rotor-stator
phase. Our aim was to provide a detailed characterization of the C$_{60}$
and cubane intermolecular dynamics in the crystal. Secondly, we have studied the polymerization process in order to identify the possible conformations of the resulting copolymer and the changes that may eventually occur on the electronic properties of the crystal.

\section{Methodology}

The atomistic simulations were performed at two distinct levels: classical and
tight-binding approximations. We have used classical molecular dynamics (MD)
to investigate molecular motions in the rotor-stator phase. The use of such type of approximation is well justified because in that phase the main interaction between C$_{60}$ and cubane molecules is due to weak van der Waals forces. Furthermore, no covalent bondings among the molecules have been observed in the temperature range covered by the rotor-stator phase. The used force-field for the atom and bond interactions was the well-known \textsc{Charmm} parameterization \cite{charmm} in the \textsc{Namd} parallel molecular dynamics code \cite{namd}. In this type of simulation we investigated a supercell composed of 3$\times$3$\times$3 fcc unit cells of the molecular crystal (108 fullerene and 108 cubane molecules). Minimization of this system resulted in a lattice parameter of 14.50 {\AA}, only 1.6 \% smaller than the experimental value (14.74 {\AA} \cite{pekker1}). The system was equilibrated during 50 ps in the canonical ensemble (NVT) followed by 200 ps runs in the microcanonical ensemble (NVE), where the molecular dynamics were analyzed. This equilibration run was sufficient for the system to reach thermodynamical equilibrium with the velocities following the Maxwell-Boltzmann distribution. We considered two temperatures (200 K and 400 K) within the range of the rotor-stator phase. Equilibration was performed carrying out Langevin dynamics with a damping coefficient 1 ps$^{-1}$ with the Newtonian equations integrated with the Br\"{u}nger-Brooks-Karplus method \cite{brunger}. In the NVE simulations the Newtonian equations of motion were numerically integrated by using the Verlet algorithm \cite{frenkel} with a timestep of 0.5 fs. 

The classical approximation using \textsc{Charmm} is no longer valid to study the polymerization process due to the existence of the possible bonding breaking/formation not possible in the \textsc{Charmm} approach. We have initially performed calculations with the well-known Tersoff-Brenner empirical potential designed for carbon systems \cite{r19}, which allows bonding breaking and formation during the evolution of the system, but it did not correctly describe the cubane geometry. So in order to investigate the polymerized phase we have described the interaction among atoms with a tight-binding model. This approach is intermediate between the classical empirical models and the fully \textit{ab initio} density functional theory. Due to the size of the system investigated here for studying the polymerization process (304 atoms in the unit cell; C$_{272}$ H$_{32}$) the tight-binding approximation presents the best compromise between accuracy and computational cost. 

We have then carried out tight-binding molecular dynamics (TBMD) simulations to investigate the polymerization process at different temperatures. We used the tight-binding model developed by Porezag \textit{et al.} \cite{dftb} implemented in the \textsc{Trocadero} program \cite{trocadero}. This model includes explicitly the non-orthogonality of the $s$-$p$ basis, in which the hopping matrix elements are obtained directly from density functional theory calculations using the same basis set but disregarding three-center contributions to the Hamiltonian. This method has been successfully applied to the prediction of allotropic forms of carbon \cite{terr1} and polymerization of C$_{60}$ inside carbon nanotubes \cite{coalesc} and it has been proven to combine accuracy and reduced computational effort, specially for large systems.

Due to the high computational cost of TBMD, a single fcc unit cell has
been simulated during 40 ps (timestep of 0.5 fs) at 2100 K, 2300 K, 2600
K, and 3100 K. These temperatures are higher than those of the
polymerization experiments, i.e, 470 K to 680 K \cite{pekker2,pekker3}.
Such high temperatures are needed in the TBMD simulations in order to accelerate the polymerization process (experimentally taking from
seconds to hours for completion) to accessible timescales
of the order of hundreds of picoseconds within current computational
capabilities. Experimental results indicate that the polymerization
reaction yields to only a slight expansion of the lattice
\cite{pekker2,pekker3}. Based on that, we considered the volume of the
system fixed during the simulations using the lattice parameter obtained
from minimization of a fcc unit cell within our tight-binding model. The
lattice parameter obtained was 15.08 {\AA}, only 2.3 \% larger than the
experimental one. We attribute this small difference to the not very accurate description of van der Waals interactions within our tight-binding model. This is not a
concern in the sense that for the high temperatures considered here,
non-bonded interactions should play a minor role. The equations of
motion were numerically integrated using the Generalized Leap-Frog
algorithm \cite{calvo} and the temperature was controlled by means of
the Nos\'{e}-Poincar\'{e} thermostat \cite{nosepoinc}. Only the $\Gamma$-point has been used for Brillouin-zone sampling. 

\section{Results and Discussions}

\subsection{I. Rotor-stator phase}

The obtained results with classical MD simulations confirm the
rotor-stator behavior of the molecular crystal, where the cubane
molecules remain basically fixed (presenting only thermal vibrations)
while C$_{60}$ fullerenes show rotational motions. As expected, we have not observed
any translational motion of any molecule in the crystal. In
Fig. 2 we present a mapping of the visited positions from a carbon atom
of the C$_{60}$ and of the cubane during 200 ps at
400 K. We can see that each atom of the cubane molecule only oscillates
around its initial equilibrium position, whereas a targeted carbon atom of C$_{60}$ walks through all the allowed spherical surface for the same period of time. The movie1 in the supplemental material provides a better visualization of these motions \cite{mov}.

The observed movement for the fullerenes presents very interesting
aspects. It is not a simple free rotation around a specific (and fixed)
axis but a composed rotation instead, i.e., the rotation axis varies continuously with time. In this sense the rotor only executes a small fraction of a full cycle until the next change in the orientation of the rotational axis. In this case, due to environment presented by the neighboring cubane molecules, each fullerene atom develops (not independently) a random walk in angle over the allowed spherical surface.

In order to better characterize the molecular dynamics of C$_{60}$ and cubane
in the rotor-stator phase we computed the second-rank single-particle time correlation function $C(\tau)$ defined as
\begin{equation}
\displaystyle C(\tau) = \frac{1}{N}\sum_{n=1}^{N}{\langle P_2[\mathbf{u}_n(t_0+\tau)\cdot\mathbf{u}_n(t_0)]\rangle_{t_0}}.
\end{equation}
Here $N$ is the number of C$_{60}$ molecules, $P_2(x)$ is the Legendre
polynomial of order 2, and $\mathbf{u}_n(t)$ is the normalized
orientational vector connecting the $n$th C$_{60}$ center of mass with
an arbitrary atom in the $n$th C$_{60}$ at time $t$. The average is
performed over different time origins $t_0$. The $C(\tau)$ time
correlation function is useful to characterize reorientational motions
in molecular systems and its Fourier transform is related to 
Raman- and light-scattering spectra \cite{berne}.

Fig. 3 depicts $C(\tau)$ for the two temperatures
considered here. For comparison, Fig. 3 also shows $C(\tau)$ computed
analytically for a spherical free rotor \cite{berne} at 200 K with the
C$_{60}$ moment of inertia. The reorientational time-correlations for fullerenes and cubanes are markedly different. The $C(\tau)$ function for the cubanes exhibits rapid damped oscillations at short times (see insert) but no long time relaxation within the simulation time, indicating that the
cubanes perform small amplitude fast librational motions (period
$\approx$ 0.2 ps) but no tumbling, consistent with the mapping pictured in Fig. 2. The fullerenes, in contrast, exhibit nearly full reorientational relaxation within this time span at both 200 and 400 K. The fullerene reorientational relaxation in the C$_{60}$-cubane crystal at these temperatures differs significantly from the free rotor behavior. The behaviors of fullerene and the free rotor coincide only at short times, up to about 0.6 ps (see insert in Fig. 3).

Overall, the $C(\tau)$ function reveals three distinct relaxation regimes for the fullerene molecules in the crystal, which resemble in many ways the
reorientational relaxation of molecular liquids. A very short-time
inertial dynamics, that lasts up to $\tau \lesssim$ 1 ps, is associated
with the fullerenes ``ballistic'' or ``free'' rotational motions
\cite{berne,willians}. The subsequent oscillatory features at 
intermediate times (1 $\lesssim \tau \lesssim$ 10 ps) stem
from small amplitude hindered rotational motions (librations) of the
fullerenes under the influence of restoring forces or torques from the
environment, largely from neighboring cubane molecules. The
post-librational, long-time behavior ($\tau \gtrsim$ 10 ps) of $C(\tau)$ is 
well described by an exponential decay characterizing a regime of rotational diffusive dynamics. Single exponential fits to $C(\tau)$ at long times yield
characteristic reorientation relaxation times of about 40 ps at 200 K
and 13 ps at 400 K for the fullerene molecules in these crystals. The
overall relaxation time, $\tau^J$, obtained from the time-integral of $C(\tau)$, $\tau^J=\int_0^{\infty} C(\tau) d\tau$, which is related to the
nuclear magnetic resonance (NMR) relaxation rate under extreme narrowing conditions, turns out 34 and 9
ps at 200 and 400 K, respectively. Our simulations suggest that the reorientational relaxation of the fullerenes is nearly one order of magnitude slower than that of a low molecular-weight nonpolar aromatic molecular crystal such as benzene \cite{benzene} (moment of inertia, $I \sim$ 14$\times$10$^{-46}$ kg.m$^2$) in this
temperature range ($\tau^J=$ 0.5 $-$ 6.0 ps), but somewhat more similar to 
that of carboranes ($I \sim$ 5$\times$10$^{-45}$ kg.m$^2$), despite the differences between the molecular rotational motions in these crystals \cite{carborane-1,carborane-2}. 

It is also interesting to compare the behavior of C$_{60}$ in the C$_{60}$-cubane crystal with the dynamics of the crystalline C$_{60}$. Johnson \textit{et al.} have investigated the rotational dynamics of C$_{60}$ in the solid state using NMR over the temperature range of 240 up to 331 K \cite{science-johnson}. They observed that the reorientational correlation time follows an Arrhenius temperature dependence with activation energy of approximately 1.4 kcal/mol for the rotor phase of crystalline C$_{60}$. The value of the activation energy is of the same order of the energy barriers found in the rotational potential maps (Fig. 5). If we consider the rotor phase in the fcc structure (temperatures above 260 K \cite{tycko}) and extrapolate  the NMR temperature dependence line of the reorientational correlation time in \cite{science-johnson} up to a temperature of 400 K, we obtain 4.6 ps \cite{correl}. The corresponding correlation time for the free gas C$_{60}$ is about 2.5 ps \cite{science-johnson}. Therefore, our simulations indicate that the reorientational relaxation time for C$_{60}$-cubane crystal is about two times as long as the expected time for the C$_{60}$ in the solid state, and almost four times as long as the estimated time for unhindered gas-phase rotation at around 400 K. 

Fig. 4 depicts the frequency spectrum of $C(\tau)$ for the two temperatures investigated here. Two peaks are evident in the spectra: one at 6 cm$^{-1}$ and another at 14.5 cm$^{-1}$. The latter is more evident in the case of 200 K while the former is more pronounced at 400 K. We associate these peaks mainly with the molecular reorientation within the librational regime.  

In order to obtain insights into the origin of
such peaks we first calculated the rotational potential energy map for the fullerene
due to van der Waals bare interactions with its six neighboring cubane
molecules (Fig. 1). For these calculations we kept the cubane molecules fixed and varied the angular orientation of the fullerene, neglecting any libration-phonon coupling. The angular orientation of the fullerene inside the cavity (Fig. 1) is characterized by the Euler angles ($0\leq\phi <2\pi$, $0\leq\theta<\pi$, and $0\leq \psi <2\pi$), and determined by the rotation matrix $\mathbf{R}(\phi,\theta,\psi)$. For each combination $(\phi,\theta,\psi)$ the potential energy $U$ was calculated leading to a four-dimensional map ($U(\phi,\theta,\psi)$). The three-dimensional potential energy map calculated for each value of $\phi$ was then used to determine local minima. For each minimum we calculated the corresponding rotational frequency $\omega_r$ through 
\begin{equation}
\displaystyle \omega_r^2= \frac{1}{2I}\frac{\partial^2 U }{\partial \xi^2} 
\end{equation}
where $I$ is the C$_{60}$ moment of inertia (9.75 $\times$ 10$^{-44}$ kg.m$^2$) and $\xi=$ $\psi$, $\theta$. These minima represent local rotational traps for the fullerenes in the crystal. We can see from Fig. 5 that different minimum valleys  appear depending upon the fullerene orientation. Fig. 6(a) shows the $\omega_r$ distribution obtained from the potential energy mapping of a C$_{60}$ fullerene surrounded by six cubane molecules (Fig. 1) for 24 different values of $\phi$. The distribution shows frequencies up to about 10 cm$^{-1}$ with a maximum around 6 cm$^{-1}$. The frequency at the maximum of the distribution corresponds to the frequency of the main spectral peaks in Fig. 4, thus showing compelling evidence that the main peaks of the reorientational frequency spectra (Fig. 4) are related to (hindered) rotational motions of the fullerenes under the potential energy wells due to the neighboring cubane molecules. We have not found any evidence of $\omega_r$ values around 14 cm$^{-1}$ in our limited search (24 $\phi$ values). However, this simplified analysis neglects any effects due to lattice phonons (and internal vibrations of fullerenes and cubanes). It is physically sensible to expect that lattice phonons act as sources of torques thus influencing the librational motions of the fullerenes. In order to examine to what extend the  librational motions of the fullerenes and lattice vibrations may be coupled, we have computed the time correlation function $C_v(\tau)$ for the fullerene center of mass velocity $\mathbf{v}_{\mathrm{cm}}(t)$:
\begin{equation}
\displaystyle C_v(\tau) = \frac{1}{N}\sum_{n=1}^{N}{\langle \mathbf{v}_{\mathrm{cm}}(t_0+\tau)\cdot\mathbf{v}_{\mathrm{cm}}(t_0)]\rangle_{t_0}},
\end{equation}
and determined its spectral density $C_v(\omega)$ using Fourier transformation. Results depicted in Fig. 6(b) show that there are indeed lattice vibrational modes in the 12-17 cm$^{-1}$ region, thus supporting our interpretation.

\subsection{II. Polymerized phases}

Having described the dynamical behavior of the rotor-stator phase, we also investigated the polymerized phases. Cubanes can thermally isomerize into different compounds. In the experiments reported by Kov\'ats \textit{et al.} \cite{pekker2a} they showed that cubane molecules inside the C$_{60}$-cubane crystals isomerize at the same temperature (about 200$^\mathrm{o}$C) as in their free standing forms. Kov\'ats \textit{et al.} also suggested that, among the most frequent decomposition products of cubane \cite{pekker2a,li} (Fig. 7), dihydropentalene (DHP) and styrene (STY) appear as the most probable compounds to form alternating copolymers with C$_{60}$. Table I presents the energy gain for the cubane isomerization into different products obtained from pyrolysis experiments \cite{li} and also from the present calculations. We can see from the Table I that STY appears as the most stable product, followed by DHP. This is also predicted by the tight-binding model used here, as well as, by more sophisticated density functional theory calculations. Furthermore, both types of calculations predict the same stability ordering. The calculations are consistent with the experimental values, except for the bicyclooctatriene (BCT) and cyclooctatetraene (COT) products which appear exchanged.

Martin \textit{et al.} experimentally investigated cubane in the temperature range 230$-$260$^\mathrm{o}$C having observed its decomposition to COT and further fragmentation to benzene (BEN) $+$ acetylene (C$_2$H$_2$) at low pressures \cite{mart1,mart2,mart3}. Experimental results also indicate that further products of COT are DHP at 700$-$850 K and STY and BEN $+$ C$_2$H$_2$ at higher temperatures \cite{li}. These results suggest that despite STY and DHP being the most stable products, the isomerization pathway has COT as an intermediate product in these processes.

In order to analyze the cubane decomposition process in its gas phase, we have carried out tight-binding molecular dynamics simulations of a system composed by eight cubane molecules during 50 ps at different temperatures. These simulations also allowed us to estimate, at least approximately, the simulation temperature which cubane molecules begin to isomerize. In the simulation time scale used here, we have verified that at 1000 K no isomerization occurs. When the temperature is increased to 1300 K, COT molecules are present, and at 1900 K COT eventually decomposes into BCT or DHP molecules. This process is consistent with experimental findings \cite{li,mart1,mart2,mart3} and may be used to map simulation and actual experimental temperature scales. Thus, a temperature of $\sim$~1900 K in the simulations can be mapped into the actual polymerization temperature of $\sim$200$^\mathrm{o}$ C. We analyzed the C$_{60}$-cubane crystal at temperatures larger than 1900 K which ensures, therefore, the decomposition of cubane molecules. 

In fact, STCO, COT, BCT, and BEN $+$ C$_2$H$_2$ were already present at 2100 K after 10 ps of TBMD simulations. Bonding between cubane products and C$_{60}$ begins to occur after 20 ps, but they are not very stable. After 40 ps, the presence of COT is still observed as well as bonding among products and C$_{60}$. Eventually bonding among C$_{60}$ molecules through products of the cubane decomposition were observed, as shown in Fig. 8 (a).

After 10 ps at 2300 K all the cubane molecules have been decomposed and DHP compounds begin to appear. In addition, we have observed hydrogen atoms bonded to the C$_{60}$ molecules. After 40 ps, covalent bonding between BEN and C$_{60}$ as well as between DHP and C$_{60}$ were observed as shown in Fig. 8 (b). When the temperature is further increased to 2600 K, all the cubane molecules were decomposed into BEN $+$ acetylene after 20 ps. The copolymerization of C$_{60}$ via cubane products is more easily observed after 30 ps as shown in Fig. 9 (a).

At 3100 K, the extension of the copolymerization is much larger than that observed at 2100 K but clearly shows signs of increased disorder. This can be seen in Fig. 9 (b) where three fullerenes are connected through fragments of cubane molecules. In this case, the crystal shows features reminiscent of an amorphous phase. We also observed that some fullerene cages were opened during the polymerization process at 3100 K. 

Based on these results, crystalline and amorphous aspects are present for 2100-2600 K while a solid with a more pronounced amorphous characteristic is observed for 3100 K. These findings agree with the experiments performed by Pekker \textit{et al.} \cite{pekker1}. The evolution of the polymerization process with the temperature can be seen in Fig. 10 through the radial distribution function $g(r)$ for 2100 K and 3100 K after 40 ps. For comparison purposes we also show $g(r)$ for the crystal structure at 0 K. The peaks at 1.39, 1.44, 2.46, and 2.85 {\AA} correspond to the first, and second nearest neighbors in C$_{60}$, in agreement with neutron-diffraction measurements at 300 K \cite{rdf-c60}. The peak at 3.56 {\AA} corresponds to third nearest neighbor in C$_{60}$, and at 1.12 and 2.35 {\AA} to the C-H and C-C bonds in the cubane molecules, respectively.

The broadening of the main peaks is observed at 2100 K and, more significantly, at 3100 K after 40 ps of molecular dynamics simulation. This indicates changes in the crystal structure due to bonding among C$_{60}$ and cubane products (2100 K) as well as bonding among C$_{60}$ themselves (3100 K). The almost complete disappearance of the peak corresponding to the C-H bonds  altogether with the broadening of peaks at 3100 K clearly show the full decomposition of the cubane molecules and the further polymerization of their decomposition products with the fullerenes.

Finally, we present the observed changes in the electronic density of states (DOS) of the C$_{60}$-cubane crystal during polymerization. Fig. 11 shows the DOS for three different temperatures. These results were confirmed with density functional theory calculations \cite{dft}. At 0 K the crystal exhibits a semiconducting behavior with a bandgap of about 1.9 eV, where the electronic states near the Fermi level are mainly due to C$_{60}$ states. The DOS is significantly different at 2100 K and 3100 K. In both cases the DOS is broader than the one obtained at 0 K. Furthermore, an increased concentration of states at the Fermi level is clearly observed. These aspects correspond to signatures of the amorphous phase and disordered characteristics presented by the resulting structure after thermal heating. The band gap of 1.9 eV can be associated to a wavelength with 654 nm and consequently to an emission/absorption of an orangish light, consistent with experimental findings of Pekker \textit{et al.} in C$_{60}$-cubane crystals without heat treatment \cite{pekker1}. In addition, the broadening observed in the calculated DOS for 2100 K and 3100 K can indicate changes in the absorption properties of the crystals and might explain the experimental observations of darker crystal colors after heat treatment \cite{pekker1}.

\section{Summary and Conclusions}
Classical and tight-binding molecular dynamics simulations were used to investigate the rotational dynamics and the polymerization processes in the C$_{60}$-cubane crystal as function of temperature. Our results show that this crystal behaves like a rotor-stator system in agreement with experimental results \cite{pekker1}. The cubane molecules work as stators and fullerenes as rotors. The fullerene molecules perform free rotations only during small time periods. For larger time intervals small amplitude hindered rotational motions and random rotations were predicted. The characteristic reorientational time of C$_{60}$ in the C$_{60}$-cubane crystal was predicted to be significantly larger than that exhibited by the C$_{60}$ crystal and by a fully free rotor. The fast components of the reorientational dynamics are characterized by two well-resolved bands in the frequency spectrum centered at 6 cm$^{-1}$ and 14.5 cm$^{-1}$, which are associated with the curvature of the potential energy surface created by cubane molecules surrounding C$_{60}$ and the coupling between C$_{60}$ hindered rotations and lattice vibrations, respectively. Random copolymerization is observed when the temperature is sufficiently high to allow cubane decomposition. The products of this decomposition initiate the polymerization process covalently connecting neighboring fullerenes. In agreement with experimental results \cite{pekker2a}, dihydropentalene appears as an important cubane product in the polymerization process. However, we have also observed that cyclooctatetraene and benzene$+$acetylene (after styrene decomposition) also contribute to the copolymerization. The polymerization causes disorder in the crystal changing the local bonding environment and the electronic structure. The density of electronic states is significantly broadened and accompanied by an increase in the concentration of states in the vicinity of the Fermi level upon copolymerization at high temperatures. We hope the present results help to interpret some unclear experimental data as well as on the design of new experiments to these structures. 

\begin{acknowledgments}
We acknowledge the financial support from the IMMP/MCT, IN/MCT, THEO-NANO, Rede de Nanotubos/CNPq, and the Brazilian agencies FAPESP, Capes, and CNPq. DSG wishes to thank Profs. B. Sundqvist and S. Pekker for helpful discussions.

\end{acknowledgments}

\clearpage
\textbf{TABLES}

\begin{table}[ht]
\caption{Energy gain (in eV) of the cubane products obtained from experiments (EXP) \cite{li} and from our tight-binding (TB) \cite{dftb} and density functional theory (DFT) \cite{dft} calculations. The experimental data were obtained from the differences in the heat of formation of the compounds. }
\vspace{0.05cm} \centering
\begin{tabular}{c c c c c}
\hline
Cubane product  &  EXP & TB  & DFT  \\
\hline
Styrene         	&  4.92  & 4.50 &   2.79 \\
\hline
Dihydropentalene	&  3.80 & 3.24 &   2.06 \\
\hline
Bicyclooctatriene	&  3.02 & 2.42 &   1.33 \\
\hline
Cyclooctatetraene	&  3.36 & 0.65 &   1.12 \\
\hline
Syn-tricyclooctadiene	&  1.34 & 0.23 &  0.29 \\
\hline
\end{tabular}
\end{table}

\clearpage
\textbf{FIGURE CAPTIONS}

Fig.1: Heteromolecular crystal structure of C$_{60}$-cubane \cite{pekker1}. Each C$_{60}$ in the crystal is surrounded by six cubane molecules. In such molecular arrangement the cubane molecules work as static bearings and the C$_{60}$ behaves as a nearly free-rotating spheroid.\\

Fig.2: (Color online) Mapping of the visited positions (smaller (blue) spheres) from a carbon atom of C$_{60}$ and cubane during a 200 ps simulation at 400 K. The initial structures of the C$_{60}$ and cubane are superimposed to the mapping in order to help the visualization.\\

Fig.3: (Color online) Single-particle correlation function $C(\tau)$
vs $\tau$ for the C$_{60}$ fullerene and cubane. For comparison purposes
the $C(\tau)$ for a free fullerene spherical rotor at 200 K is also
presented. The insert shows details of $C(\tau)$ short-dynamics in a semi-log scale. The correlation function for cubane in the inset graph was expanded five times to facilitate visualization.\\

Fig.4: $\omega^2$ times the frequency spectrum of
$C(\tau)$ obtained from the cosine Fourier transform ($C(\omega)$). As
depicted, the frequency spectra would correspond roughly to the intermolecular
components of the Raman or depolarized light scattering spectra of
fullerenes in the crystal.\\

Fig.5: (Color online) Rotational potential experienced by a C$_{60}$ fullerene when surrounded by six cubane molecules (Fig. 1) as a function of the $\theta$ and $\psi$ angles for (a) $\phi=0$, (b) 30$^\mathrm{o}$, (c) 90$^\mathrm{o}$, and (d) 150$^\mathrm{o}$.\\

Fig.6: (a) Distribution of the rotational frequency $\omega_r$ obtained from the curvature of the minima on the three-dimensional potential energy maps generated by rotating a C$_{60}$ fullerene surrounded by six cubane molecules around a fixed center of mass position (cf. Fig. 1). (b) Low frequency portions of the spectral density $C_v(\omega)$ of $C_v(\tau)$ obtained from Fourier transforms.\\

Fig.7: (a) Cubane and its most frequent decomposition products: (b) syn-tricyclooctadiene (STCO), (c) cyclooctatetraene (COT), (d) dihydropentalene (DHP), (e) bicyclooctatriene (BCT), and (f) styrene (STY).\\

Fig.8: Snapshots of the tight-binding molecular dynamics simulations showing fragments of the C$_{60}$-cubane crystal at (a) 2100 K and; (b) 2300 K, after 40 ps.\\

Fig.9: Snapshots of the tight-binding molecular dynamics simulations showing fragments of the C$_{60}$-cubane crystal at (a) 2600 K and; (b) 3100 K, after 30 ps.\\

Fig.10: (Color online) Radial distribution function $g(r)$ for the C$_{60}$-cubane crystal at different temperatures.\\

Fig.11:Electronic density of states (DOS) for three different configurations of the C$_{60}$-cubane crystal. The vertical dashed line indicates the position of the  Fermi level. Non-zero density of states at the Fermi level for 0 K is due to peak broadening.

\clearpage
\textbf{FIGURES}
\vspace{1cm}

\begin{figure}[ht]
\begin{center}
\includegraphics[angle=0,scale=0.5]{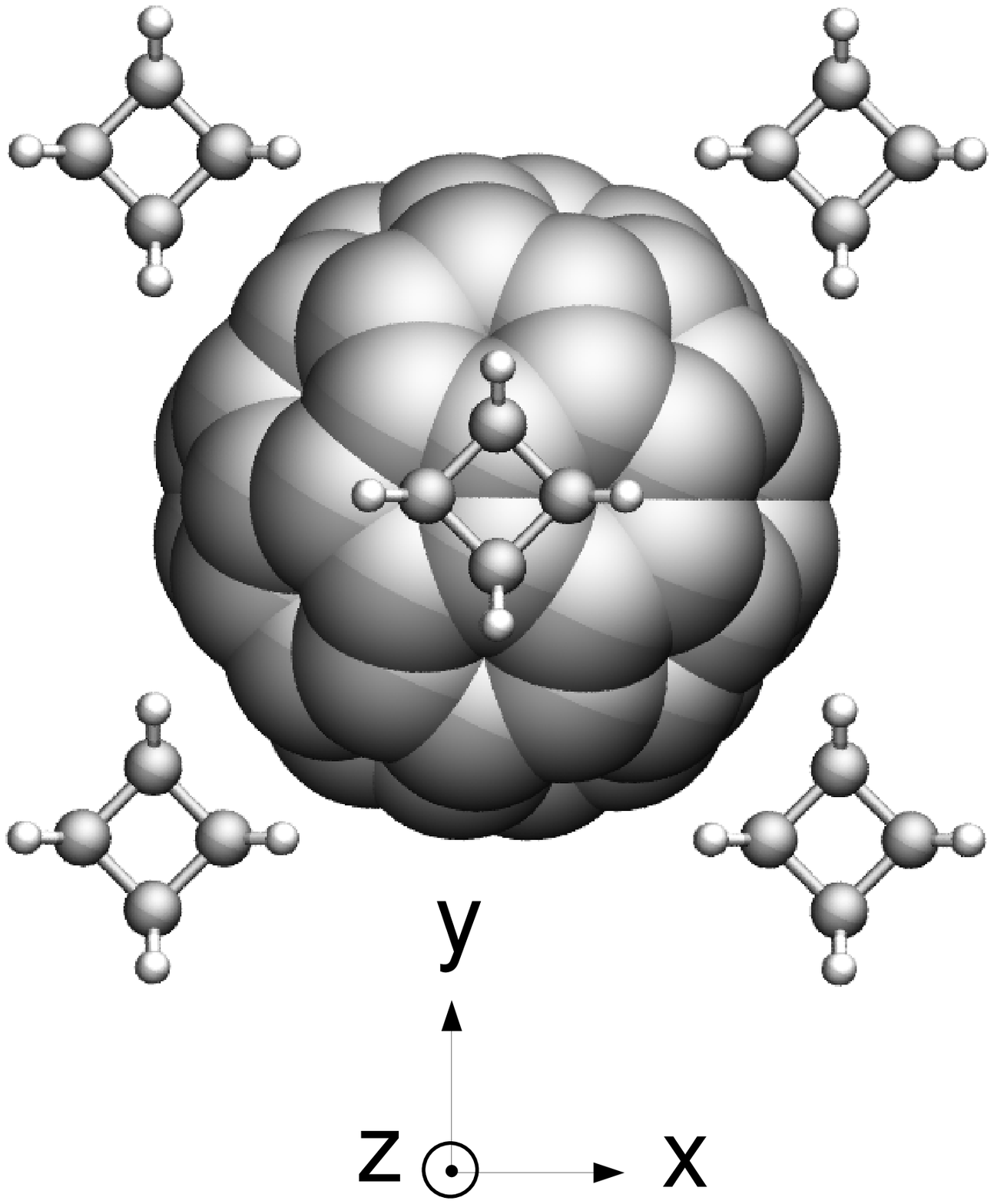}
\caption{}
\end{center}
\end{figure}

\clearpage
\begin{figure}[ht]
\begin{center}
\includegraphics[angle=90,scale=0.5]{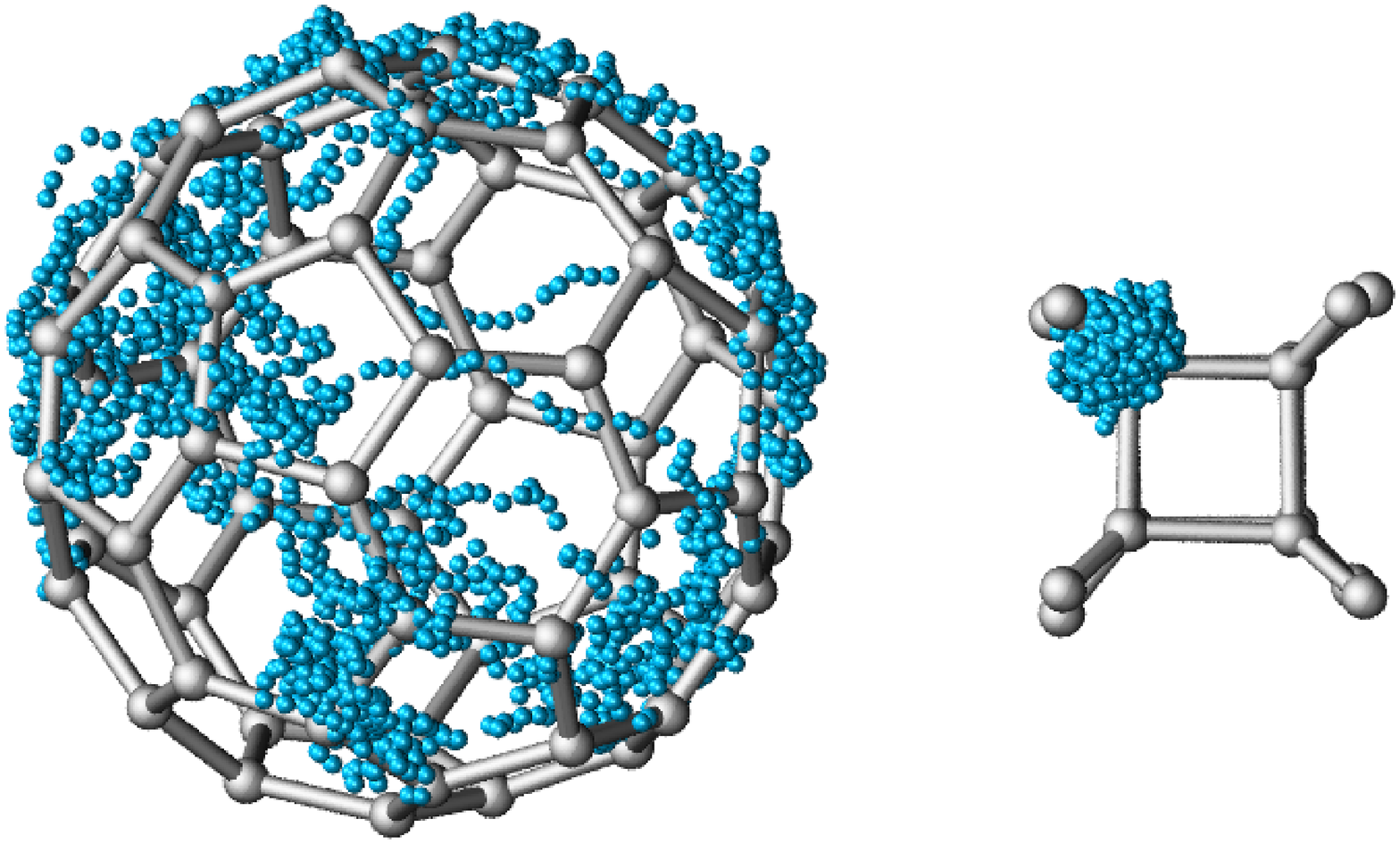}
\caption{}
\end{center}
\end{figure}

\clearpage
\begin{figure}[h]
\begin{center}
\includegraphics[angle=0,scale=0.6]{fig-3.eps}
\caption{}
\end{center}
\end{figure}

\clearpage
\begin{figure}[h]
\begin{center}
\includegraphics[angle=0,scale=0.6]{fig-4.eps}
\caption{}
\end{center}
\end{figure}

\clearpage
\begin{figure}[h]
\begin{center}
\includegraphics[angle=0,scale=0.6]{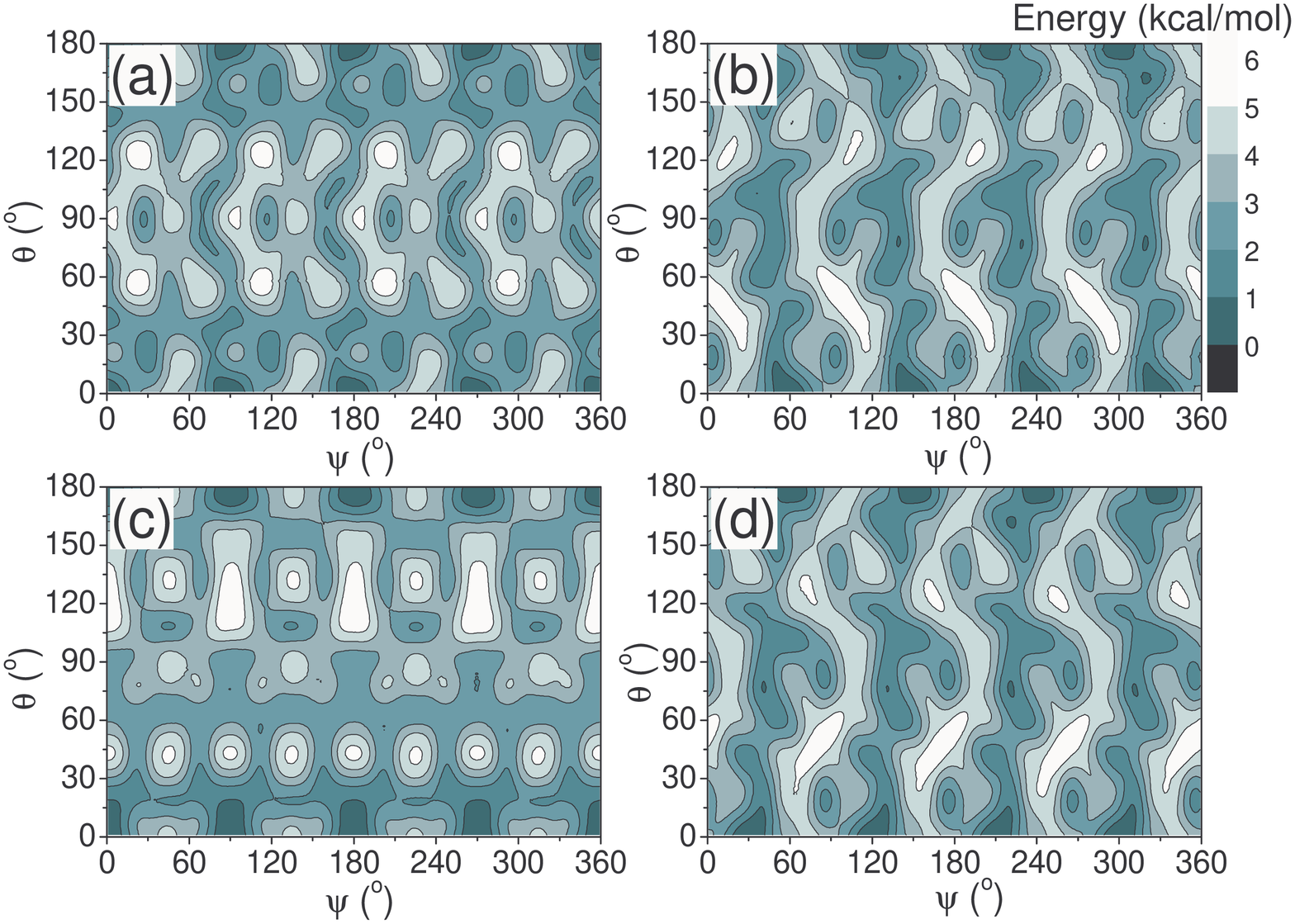}
\caption{}
\end{center}
\end{figure}

\clearpage
\begin{figure}[h]
\begin{center}
\includegraphics[angle=0,scale=0.6]{fig-6.eps}
\caption{}
\end{center}
\end{figure}

\clearpage
\begin{figure}[h]
\begin{center}
\includegraphics[angle=0,scale=0.6]{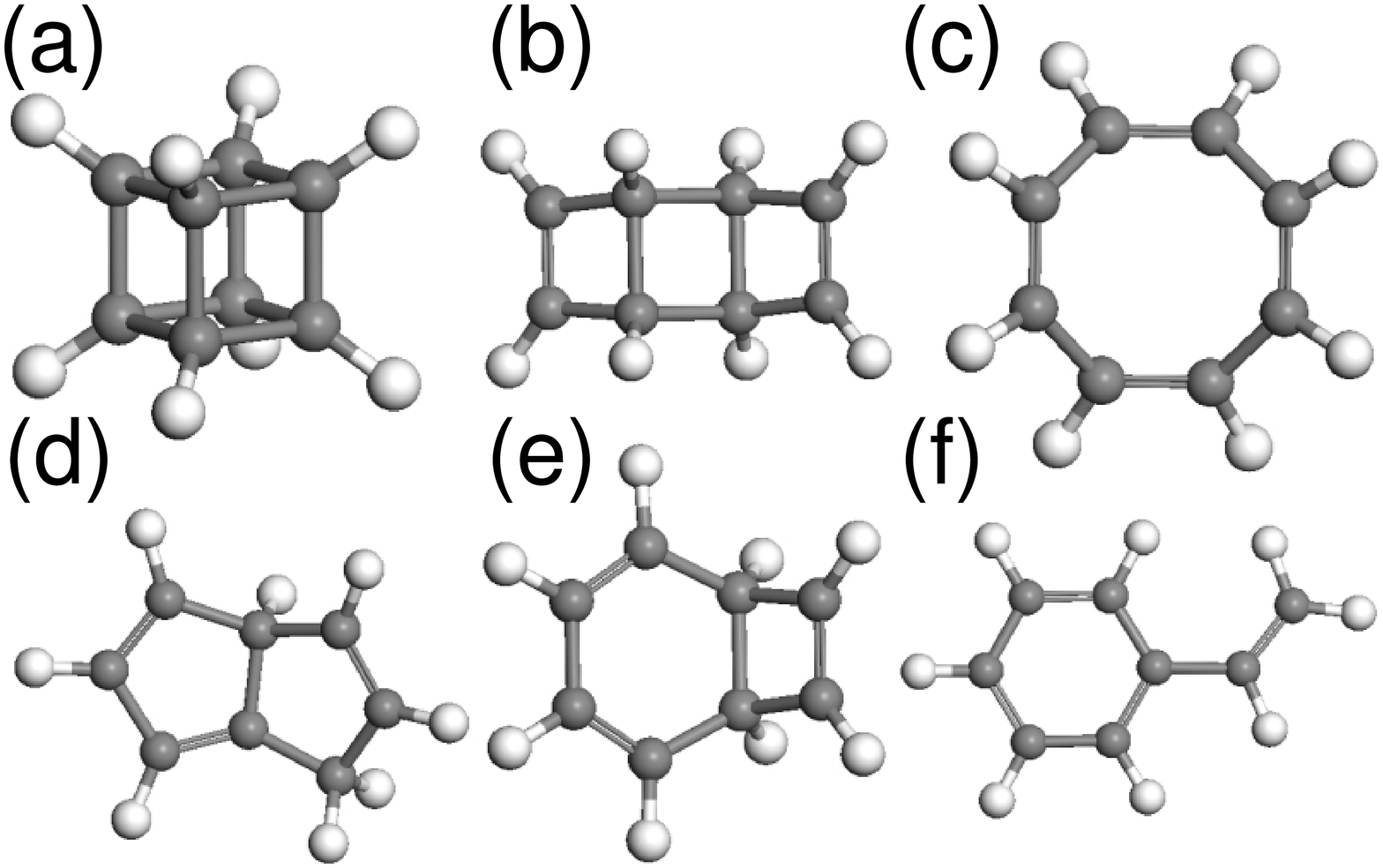}
\caption{}
\end{center}
\end{figure}

\clearpage
\begin{figure}[h]
\begin{center}
\includegraphics[angle=0,scale=0.7]{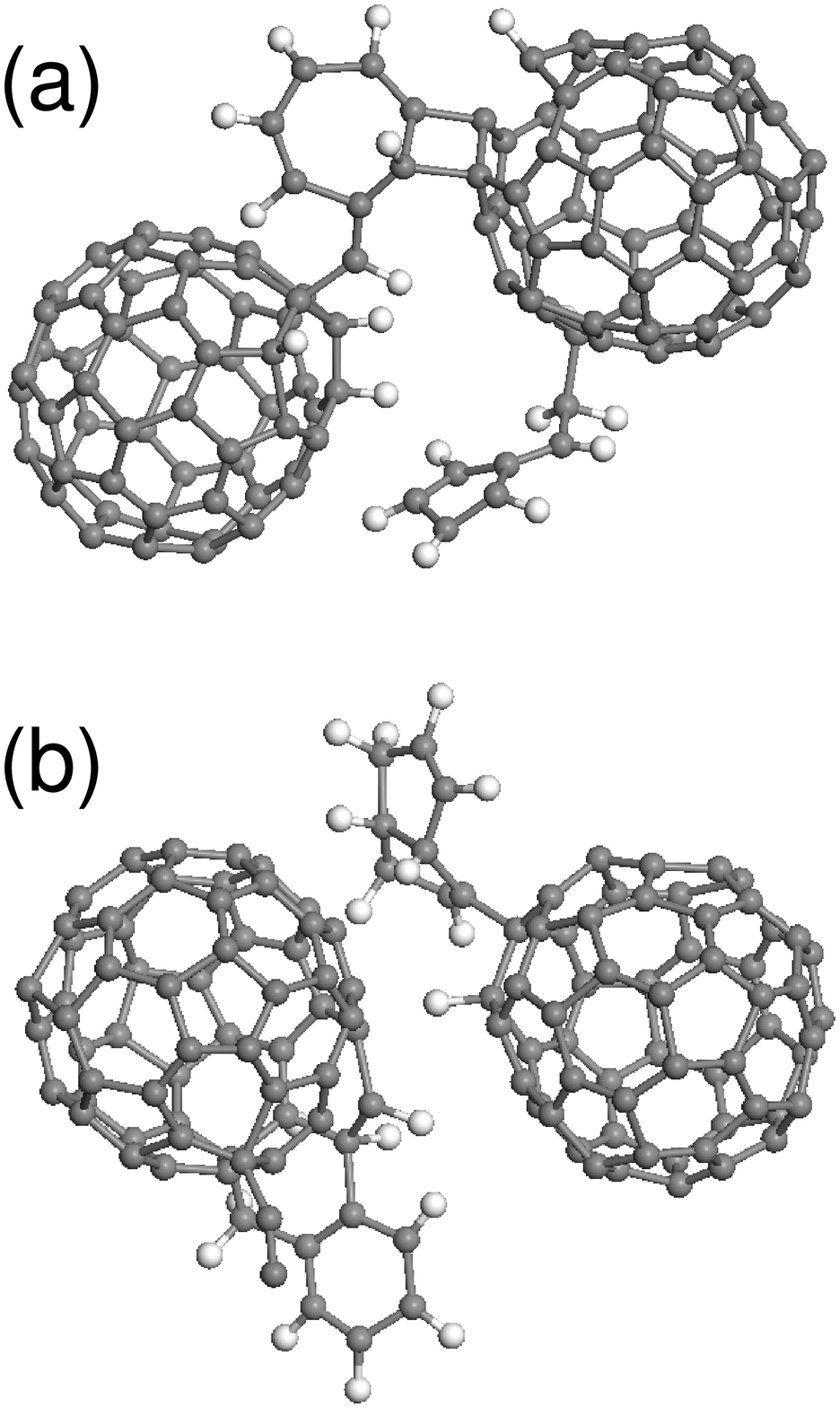}
\caption{}
\end{center}
\end{figure}

\clearpage
\begin{figure}[h]
\begin{center}
\includegraphics[angle=0,scale=0.7]{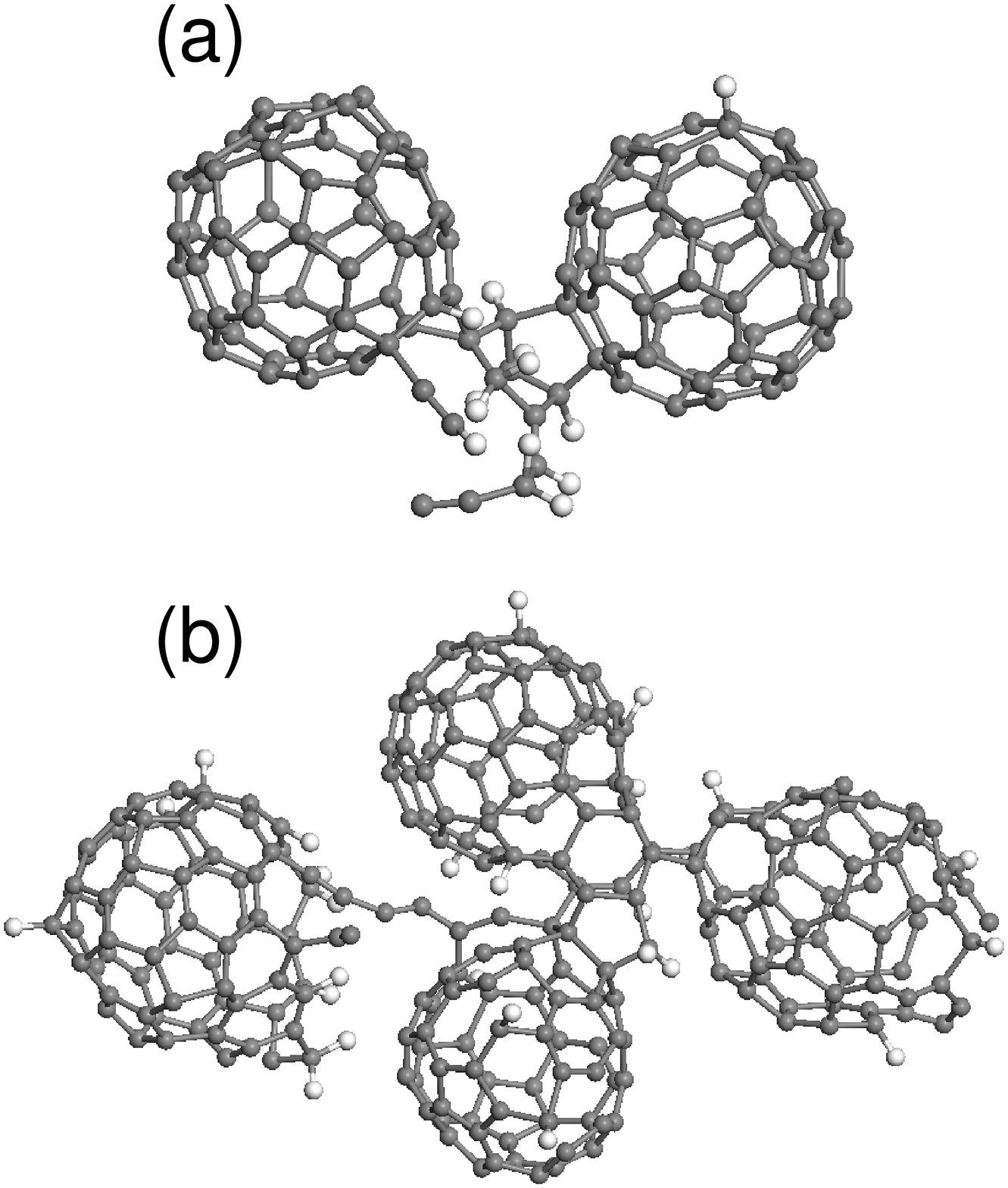}
\caption{}
\end{center}
\end{figure}

\clearpage
\begin{figure}[h]
\begin{center}
\includegraphics[angle=0,scale=0.6]{fig-10.eps}
\caption{}
\end{center}
\end{figure}

\clearpage
\begin{figure}[h]
\begin{center}
\includegraphics[angle=0,scale=0.7]{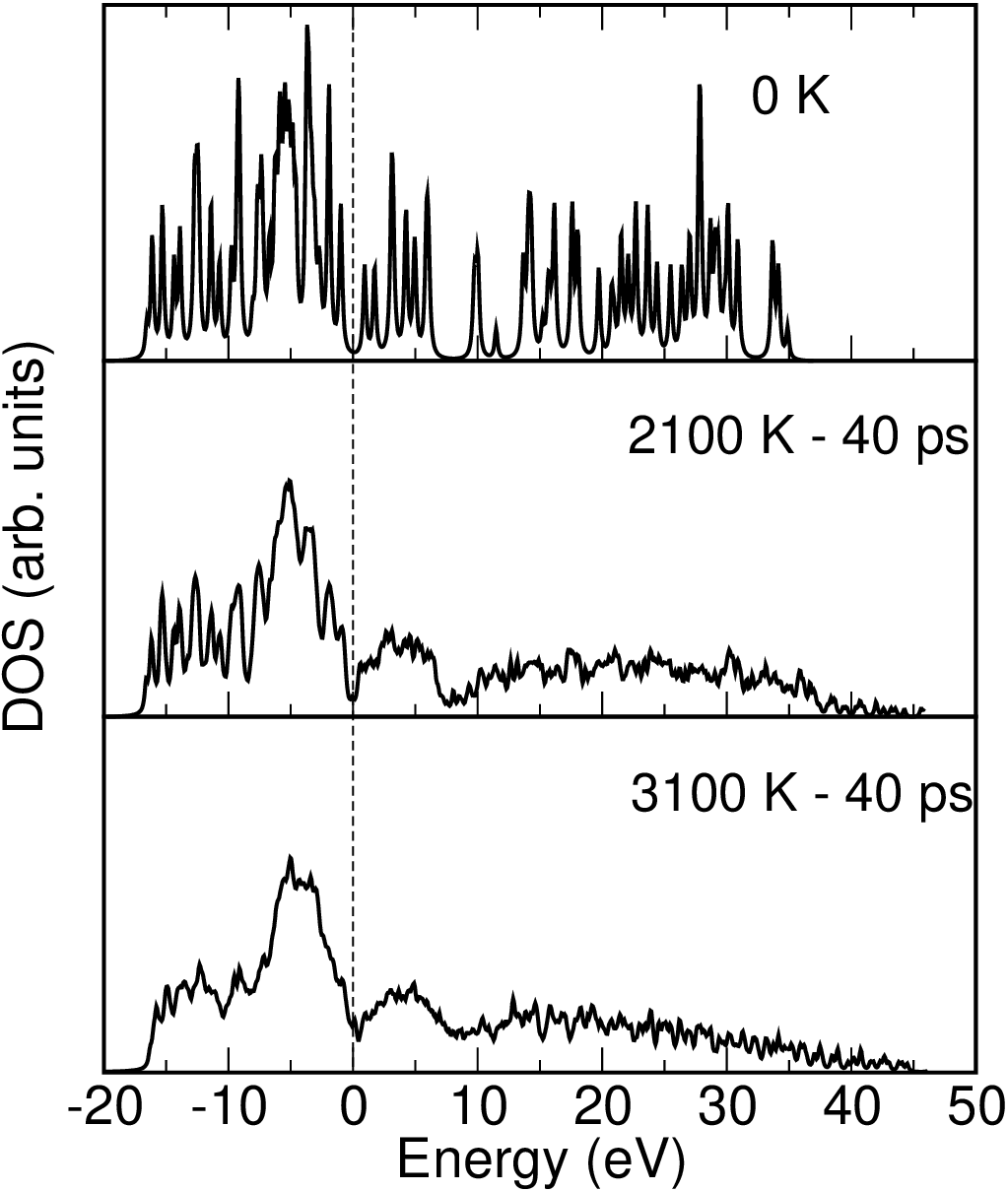}
\caption{}
\end{center}
\end{figure}


\begin{thebibliography}{99}
\bibitem{1} H. W. Kroto, J. R. Heath, S. C. O'Brien, R. F. Curl, R. E. Smalley, \textit{Nature}  {\bf 318}, 162 (1999). 

\bibitem{livro} M. S. Dresselhaus, G. Dresselhaus, P. C. Eklund, Science of Fullerenes and Carbon Nanotubes, Academic Press, San Diego, CA, 1995.

\bibitem{cox}P. A. Heiney, J. E. Fischer, A. R. McGhie, W. J. Romanow, A. M. Denenstein, J. P. McCauley, Jr., A. B. Smith, D. E. Cox, \textit{Phys. Rev.
Lett.} {\bf 66}, 2911 (1991).


\bibitem{saito}S. Saito, A. Oshiyama, \textit{Phys. Rev. Lett.} \textbf{66}, 2637 (1991).

\bibitem{sun}B. Sundqvist, \textit{Adv. Phys.} \textbf{48}, 1 (1999).

\bibitem{iwasa}Y. Iwasa, T. Arima, R. M. Fleming, T. Siegrist, O. Zhou, R. C. Haddon, L. J. Rothberg, K. B. Lyons, H. L. Carter, Jr., A. F. Hebard, R. Tycko, G. Dabbagh, J. J. Krajewski, G. A. Thomas, T. Yagi, \textit{Science} \textbf{264}, 1570 (1994).

\bibitem{regueiro}M. N\'u\~nez-Regueiro, L. Marques, J -L. Hodeau, O. B\'ethoux, M. Perroux, \textit{Phys. Rev. Lett.} \textbf{74}, 278 (1995).

\bibitem{Goze} C. Goze, F. Rachdi, L. Hajji, M. N\'u\~nez-Regueiro, L. Marques, J -L. Hodeau, M. Mehring, \textit{Phys. Rev. B} \textbf{54}, R3676 (1996).

\bibitem{Jansson} A. V. Talyzin, L. S. Dubrovinsky, T. Le Bihan, U. Jansson, \textit{Phys. Rev. B} \textbf{65}, 245413 (2002).

\bibitem{Rao} A. M. Rao, P. Zhou, K. A. Wang, G. T. Hager, J. M. Holden, U. Wang, W. T. Lee, X. X. Bi, P. C. Eklund, D. S. Cornett, M. A. Duncan, I. J. Amster, \textit{Science} \textbf{259}, 955 (1993).

\bibitem{Ranjan} K. Ranjan, K. Dharamvir, V. K. Jindal, \textit{Physica B} \textbf{371}, 232 (2006).

\bibitem{Stenger} C. H. Pennington, V. A. Stenger, \textit{Rev. Mod. Phys.} \textbf{68}, 855 (1996).

\bibitem{JPCS} D. Varshneya, M. Varshneya, R. K. Singhb, R. Mishraa, \textit{J. Phys. Chem. Sol.} \textbf{60}, 579 (1999).

\bibitem{gas1} A. N. Aleksandrovskii, V. G. Gavrilko, V. B. Eselson, V. G. Manzhelii, B. G. Udovidchenko, V. P. Maletskiy, B. Sundqvist, \textit{Low Temp. Phys.} \textbf{27}, 1033 (2001).

\bibitem{gas2} T. B. Tang, M. Gu \textit{Phys. Sol. Stat.} \textbf{44}, 631(2002).

\bibitem{Russian} Yu. M. Shulga, V. M. Martynenko, A. F. Shestakov, S. A. Baskakov, S. V. Kulikov, V. N. Vasilets, T. L. Makarova, Yu. G. Morozov, \textit{Russ. Chem. Bull. Int. Ed.} \textbf{55}, 687 (2006).

\bibitem{cubano} P. E. Eaton, \textit{Angew. Chem. Int. Ed. Engl.}  {\bf 31}, 1421 (1992).

\bibitem{pekker1} S. Pekker, \'E. Kov\'ats, G. Oszl\'anyi, Gy. B\'enyei, G. Klupp, G. Bortel, I. Jalsovszky, E. Jakab, F. Borondics,  K. Kamar\'as, M. Bokor, G. Kriza, K. Tompa, G. Faigel, \textit{Nature Materials} \textbf{4} 764 (2005).

\bibitem{pekker2a} \'E. Kov\'ats, G. Klupp, E. Jakab, \'A. Pekker, K. Kamar\'as, I. Jalsovszky, S. Pekker, \textit{Phys. Stat. Sol. (b)} \textbf{243} 2985 (2006).

\bibitem{pekker2} S. Pekker, \'E. Kov\'ats, G. Oszl\'anyi, Gy. B\'enyei, G. Klupp, G. Bortel, I. Jalsovszky, E. Jakab, F. Borondics,  K. Kamar\'as, G. Faigel, \textit{Phys. Stat. Sol. (b)} \textbf{243} 3032 (2006).

\bibitem{pekker3} A. Iwasiewicz-Wabnig, B. Sundqvist, \'E. Kov\'ats, I. Jalsovszky, S. Pekker, \textit{Phys. Rev. B} \textbf{75} 024114 (2007).

\bibitem{charmm} A. D. MacKerell, Jr., D. Bashford, M. Bellott, R. L. Dunbrack Jr., J. D. Evanseck, M. J. Field, S. Fischer, J. Gao, H. Guo, S. Ha, D. Joseph-McCarthy, L. Kuchnir, K. Kuczera, F. T. K. Lau, C. Mattos, S. Michnick, T. Ngo, D. T. Nguyen, B. Prodhom, W. E. Reiher III,
B. Roux, M. Schlenkrich, J. C. Smith, R. Stote, J. Straub, M. Watanabe, J. Wiorkiewicz-Kuczera, D. Yin, M. Karplus, \textit{J. Phys. Chem. B}, {\bf 102} 3586 (1998).

\bibitem{namd} J. C. Phillips, R. Braun, W. Wang, J. Gumbart, E. Tajkhorshid, E. Villa, C. Chipot, R. D. Skeel, L. Kale, K. Schulten, \textit{J. Comput. Chem.}  {\bf 26}, 1781 (2005).

\bibitem{brunger} A. Br\"{u}nger, C. B. Brooks, M. Karplus, \textit{Chem. Phys. Lett.}  {\bf 105}, 495 (1984).

\bibitem{frenkel} D. Frenkel and B. Smit, Understanding molecular simulation: from algorithms to applications, Academic Press, San Diego, CA, 2002.

\bibitem{r19} D. W. Brenner, \textit{Phys. Rev. B} \textbf{42}, 9458 (1990).

\bibitem{dftb} D. Porezag, T. Frauenheim, T. Kohler, G. Seifert, R. Kaschner, \textit{Phys. Rev. B} \textbf{51}, 12947 (1995).

\bibitem{trocadero} R. Rurali, E. Hernandez, \textit{Comput. Mat. Sci.} \textbf{28}, 85 (2003).

\bibitem{terr1} H. Terrones, M. Terrones, E. Hernandez, N. Grobert, J. C. Charlier, P. M. Ajayan, \textit{Phys. Rev. Lett.} \textbf{84}, 1716 (2000).

\bibitem{coalesc} E. Hernandez, V. Meunier, B. W. Smith, R. Rurali, H. Terrones, M. B. Nardelii, M. Terrones, D. E. Luzzi, J.-C. Charlier,
\textit {Nano Lett.} \textbf{3}, 1037 (2003). 

\bibitem{calvo} J. M. Sanz-Serna, M. P. Calvo, Numerical Hamiltonian Problems, Chapman and Hall, New York, 1995.

\bibitem{nosepoinc} S. D. Bond, B. J. Leimkuhler, B. B. Laird, \textit{J. Comput. Phys.} \textbf{151}, 114 (1999).

\bibitem{mov}
See EPAPS Document No. XXXXXXXX for the movie mentioned in the text. 
This document may be retrieved via the EPAPS homepage
(http://www.aip.org/pubservs/epaps.html) or from ftp.aip.org in
the directory /epaps/. See the EPAPS homepage for more information.


\bibitem{berne} B. J. Berne and R. Pecora, Dynamic Light Scattering: with applications to Chemistry, Biology, and Physics, Dover Publications, Inc. Mineola, New York, 2000.

\bibitem{willians} G. Williams, \textit{Chem. Soc. Rev.} \textbf{7}, 89 (1978).

\bibitem{benzene}R. Witt, L. Sturz, A. Dolle, F. M.-Plathe, \textit{J. Phys. Chem. A} \textbf{104}, 5716 (2000).

\bibitem{carborane-1}Z. Gamba, B. M. Powell, \textit{J. Chem. Phys.} \textbf{105}, 2436 (1996).

\bibitem{carborane-2}M. Winterlich, R. B\"{o}hmer, G. Diezemann, H. Zimmermann, \textit{J. Chem. Phys.} \textbf{123}, 094504 (2005).

\bibitem{science-johnson}R. D. Johnson, C. S. Yannoni, H. C. Dorn, J. R. Salem, D. S. Bethune, \textit{Science} \textbf{255}, 1235 (1992).

\bibitem{tycko}R. Tycko, G. Dabbagh, R. M. Fleming, R. C. Haddon, A. V. Makhija, S. M. Zahurak, \textit{Phys. Rev. Lett.} \textbf{67}, 1886 (1991).


\bibitem{correl} The fitting for the range of 241 to 331 K for the reorientational correlation time $\tau_r$ in ps is $\tau_r=$ 0.81 $\exp(695/T)$ where $T$ is the temperature in K \cite{science-johnson}.

\bibitem{li} Z. Li, S. L. Anderson, \textit{J. Phys. Chem. A} \textbf{107}, 1162 (2003); and references therein.

\bibitem{dft} Density functional theory calculations were performed with the \textsc{siesta} code \cite{siesta1,siesta2} in the local density approximation based on the Perdew-Zunger construction \cite{ca} with the pseudopotential generated according to the Troullier-Martins scheme \cite{troullier}. The standard double zeta plus polarization basis was used. Both pseudopotential and basis set were optimized according to Junqueira \textit{et al.} \cite{c60basis}. 

\bibitem{siesta1} D. Sanchez-Portal, P. Ordej\'on, E. Artacho, J. M. Soler, \textit{Int. J. Quantum Chem.} \textbf{65}, 453 (1997).

\bibitem{siesta2} J. Soler, E. Artacho, J. D. Gale, A. Garc\'ia, J. Junquera,
P. Ordej\'on, D. Sanchez-Portal, \textit{J. Phys.: Condens. Matter} \textbf{14}, 2745 (2002).

\bibitem{ca} J. P. Perdew, A. Zunger, \textit{Phys. Rev. B} \textbf{23}, 5048 (1981).

\bibitem{troullier} N. Troullier, J. L. Martins, \textit{Phys. Rev. B} \textbf{43}, 1993 (1991).

\bibitem{c60basis} J. Junquera, \'O. Paz, D. S\'anchez-Portal, E. Artacho,  \textit{Phys. Rev. B} \textbf{64}, 235111 (2001).

\bibitem{mart1} H. D. Martin, T. Urbanek, P. Pfohler, R.  Walsh, \textit{J. Chem. Soc. Chem. Commun.} 964 (1985).

\bibitem{mart2} H. D. Martin, T. Urbanek, R. Walsh, \textit{J. Am. Chem. Soc.} \textbf{107}, 5532 (1985).

\bibitem{mart3} H. D. Martin, P. Pfohler, T. Urbanek, R. Walsh, \textit{Chem. Ber.} \textbf{116}, 1415 (1983).

\bibitem{rdf-c60} F. Li, D. Ramage, J. S. Lannin, J. Conceicao, \textit{Phys. Rev. B} \textbf{44}, 13167 (1991).
 
\end{thebibliography}
\end{document}